\documentclass[12pt]{iopart}
\usepackage{graphicx}
\usepackage{wrapfig}
\usepackage{xspace}

\newcommand{\AVG}[1]{\ensuremath{\left\langle #1 \right\rangle}}
\newcommand{\GeVc}{GeV\kern-0.2em /\kern-0.15em \textit{c}\xspace}
\newcommand{\MeVc}{MeV\kern-0.2em /\kern-0.15em \textit{c}\xspace}
\newcommand{\A}{\textit{A}}
\newcommand{\AGeVc}{\A\kern+0.3em \GeVc}

\newcommand{\eV}{\ensuremath{\mbox{e\kern-0.1em V}}\xspace}
\newcommand{\GeV}{\ensuremath{\mbox{Ge\kern-0.1em V}}\xspace}
\newcommand{\MeV}{\ensuremath{\mbox{Me\kern-0.1em V}}\xspace}

\begin{document}

\title[Search for the QCD critical point by NA61/SHINE at the CERN SPS]{Search for the QCD critical point by NA61/SHINE at the CERN SPS}

\author{Haradhan Adhikary for the NA61/SHINE Collaboration}

\address{Jan Kochanowski University, Kielce, Poland}
\ead{haradhan.adhikary@cern.ch}
\vspace{10pt}
\begin{indented}
\item[Contribution to: ICNFP2021 ]
\end{indented}


\begin{abstract}
The existence and location of the QCD critical point is an object of both experimental and theoretical studies. The comprehensive data collected by the NA61/SHINE during a two-dimensional scan in beam momentum (13\A-150\AGeVc) and system size (\textit{p}+\textit{p}, \textit{p}+Pb, Be+Be, Ar+Sc, Xe+La, Pb+Pb) allows for a systematic search for the critical point -- a search for a non-monotonic dependence of various correlation and fluctuation observables on collision energy and size of colliding nuclei. In particular, fluctuations of particle number in transverse momentum space are studied. They are quantified by measuring the scaling behavior of factorial moments of multiplicity distributions.\\
This contribution reviews ongoing NA61/SHINE studies to search for the critical point of the strongly interacting matter.
\end{abstract}

\vspace{0.5pc}
\hspace{3.7pc}

\section{Introduction}

\subsection{NA61/SHINE at CERN SPS}
NA61/SHINE (SPS Heavy Ion and Neutrino Experiment)~\cite{NA61} is a multi-purpose facility measuring hadron production in hadron-proton, hadron-nucleus and nucleus-nucleus collisions. It is a fixed target experiment located at the H2 beam-line in the North Area of the CERN Super Proton Synchrotron (SPS).
The main physics goals of the experiment are to study the onset of deconfinement and search for the critical point.


\subsection{NA61/SHINE detector}

\begin{figure}
    \centering
    \includegraphics[width=5.5in]{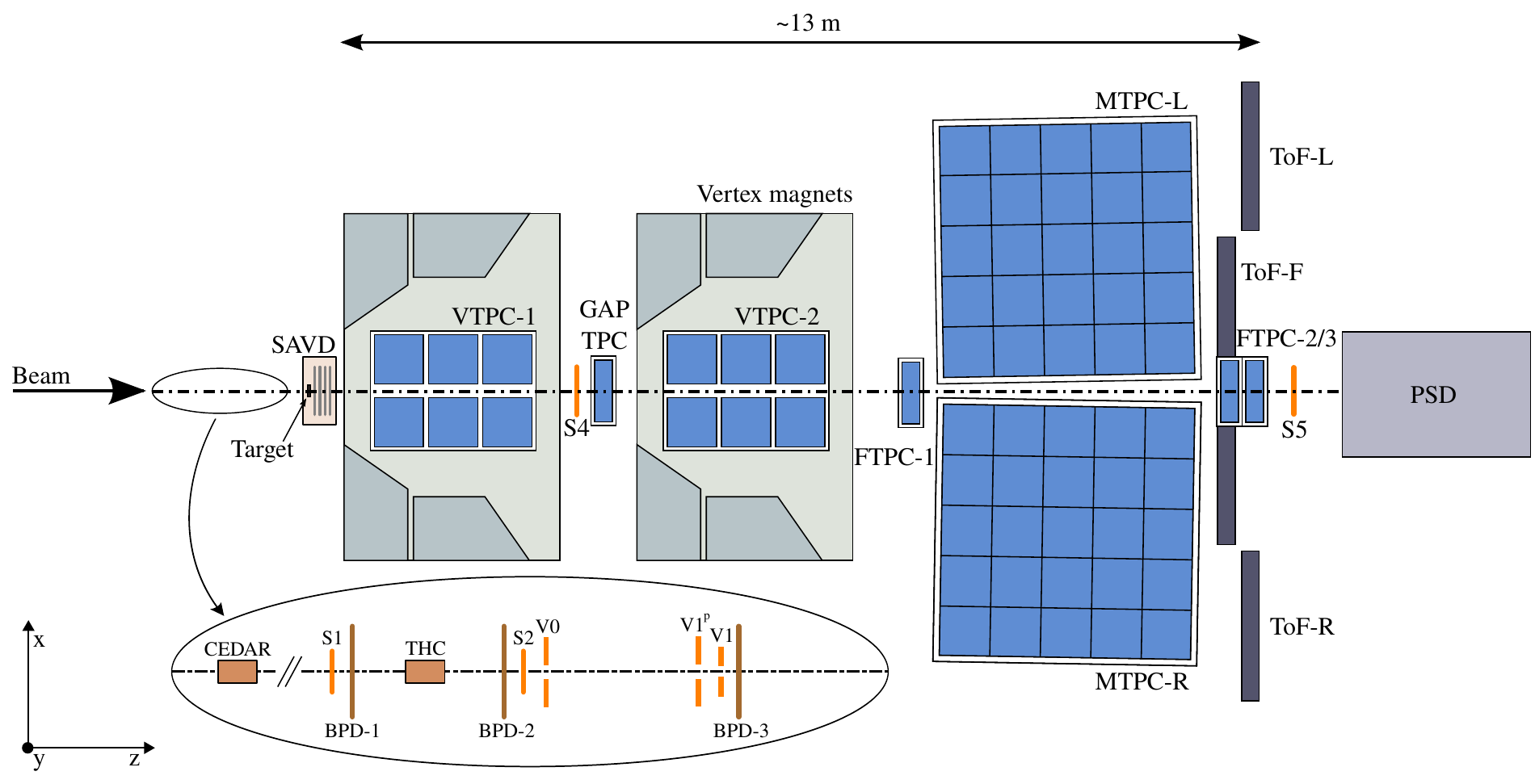}
    \caption{Schematic layout of the NA61/SHINE detection system~\cite{NA61}.
    \label{na61setup}}
\end{figure}

The NA61/SHINE detection system~\cite{NA61}, presented in Fig.~\ref{na61setup}, is a large acceptance hadron spectrometer with excellent capabilities in measuring charged particles. The main tracking devices are eight Time Projection Chambers, two of which are placed inside superconducting magnets, complemented by three Time-of-Flight detectors. This setup allows for precise momentum reconstruction and identification of charged particles.
The high-resolution forward hadron calorimeter, the Projectile Spectator Detector (PSD), is used to determine the centrality of the collision by the measurements of the forward energy (related to the number of projectile spectators, i.e., non-interacting projectile nucleons). The beam is monitored by beam detectors, which allows us to identify beam particles and precisely measure their trajectories.


\subsection{NA61/SHINE heavy-ion program}
\begin{wrapfigure}{r}{.5\textwidth}
\centering
\vspace{-\intextsep}
\includegraphics[width=3in]{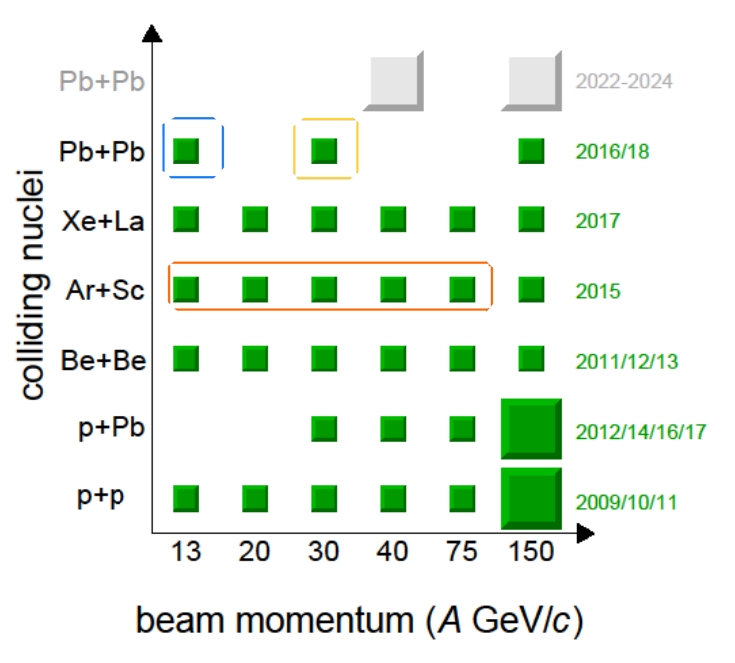}
\caption{The NA61/SHINE beam energy and system size scan }
\label{fig:beam_energy_scan}
\end{wrapfigure}

NA61/SHINE performed a two-dimensional scan (Fig.~\ref{fig:beam_energy_scan}) in collision energy (13\A-~150\AGeVc) and system size (p+p, Be+Be, Ar+Sc, Xe+La, Pb+Pb) to study the phase diagram of strongly interacting matter. The main goals of NA61/SHINE are to search for the critical point and study the onset of deconfinement. Currently, NA61/SHINE pursues the physics program focused on the open charm production measurements in order to determine the mechanism and the impact of the onset of deconfinement on it and investigate the effect of the QGP formation on J/$\psi$ production. NA61/SHINE also pursuing physics program including measurements neutrino, and cosmic ray physics.

\section{Critical point search strategies}  

\subsection{QCD critical point}

\begin{wrapfigure}{r}{.5\textwidth}
  \centering
  \vspace{-\intextsep}
  \includegraphics[width=.35\textwidth]{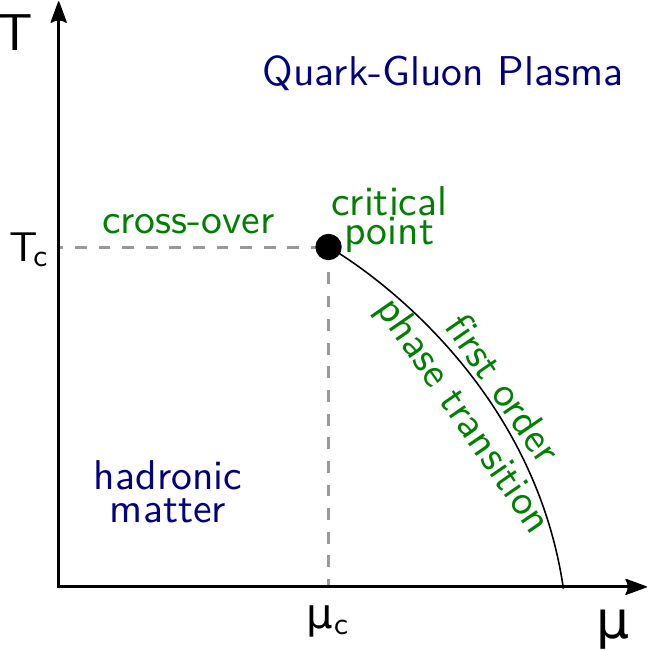}
  \caption{
    A sketch of the phase diagram of strongly interacting matter.
  }
  \label{fig:phase-diagram}
\end{wrapfigure}

A sketch of the most popular phase diagram of the strongly interacting matter is shown in Fig.~\ref{fig:phase-diagram}. At low temperature (T) and baryon chemical potential ($\mu$) the matter consists of quarks and gluons confined inside hadrons. At higher temperatures and/or baryon chemical potential, quarks and gluons may act like quasi-free particles, forming a different state of matter -- Quark-Gluon Plasma (QGP). A first-order phase transition is expected between the two phases at high $\mu$.
Critical point (CP) is a hypothetical endpoint of this first-order phase transition line that has properties of second-order phase transition~\cite{Asakawa:1989bq,Barducci:1989wi}.

\begin{figure}[!htb]
  \centering
  \includegraphics[width=0.82\textwidth]{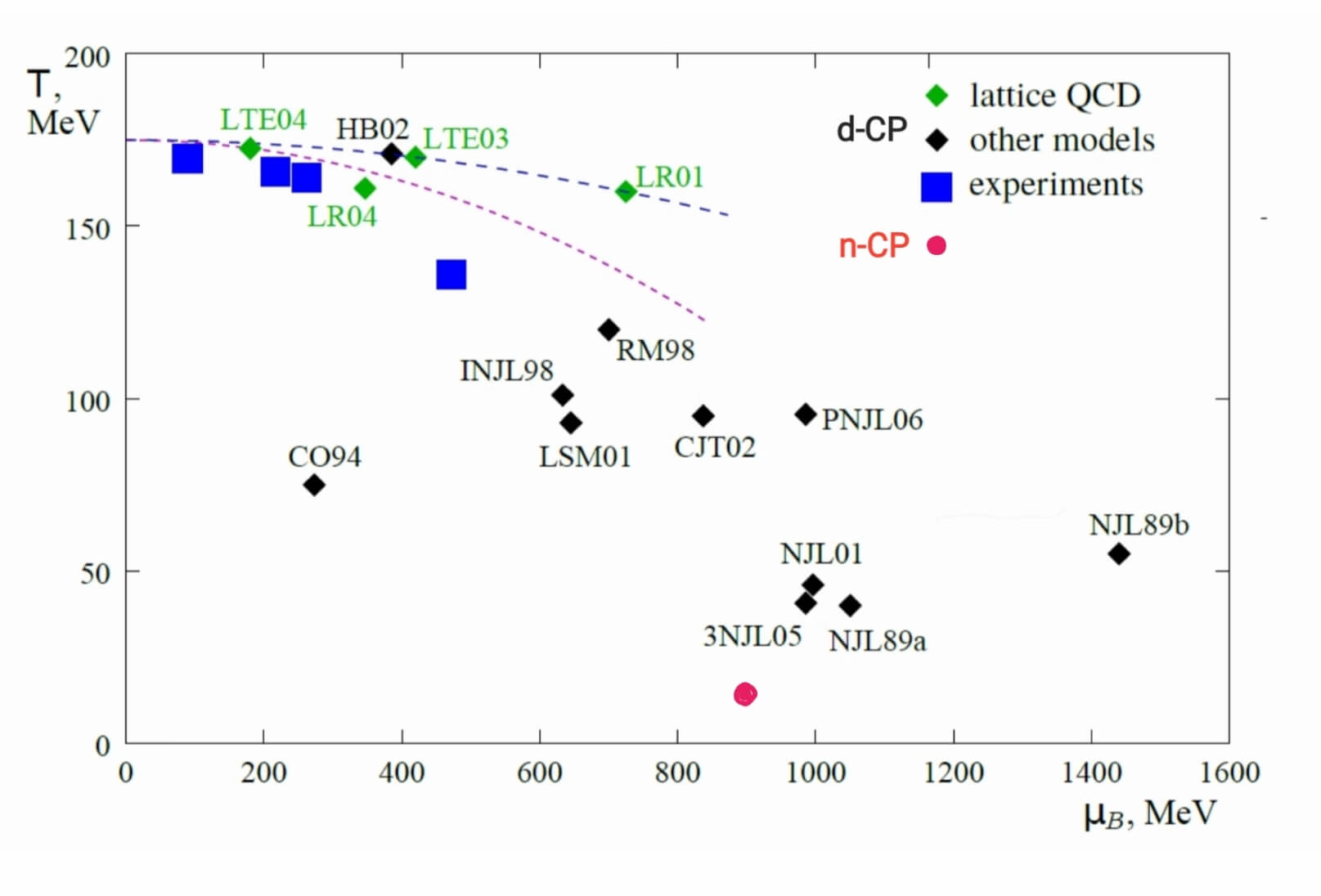}
  \caption{
    Compilation of theoretical predictions~\cite{Stephanov:2007fk} and experimental hints~\cite{Czopowicz:2020twk} on the location of the deconfinement critical
point, d-CP, in the phase diagram of strongly interacting matter. The position 
of the nuclear critical point, n-CP, as suggested by theoretical and experimental results, is indicated for comparison.
  }
  \label{fig:summary}
\end{figure}

It is expected that the QCD critical point should lead to an anomaly in fluctuations in a narrow domain of the phase diagram. However, predictions on the CP existence (Fig.~\ref{fig:summary}), its location and what and how should fluctuate are model-dependent~\cite{Stephanov:2007fk}.

\subsection{Exploring the phase diagram with heavy-ion collisions}

\begin{figure}[!htb]
  \centering
  \includegraphics[width=0.44\textwidth]{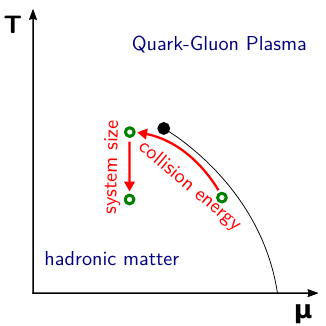}
  \vspace{3pt}\includegraphics[width=0.54\textwidth]{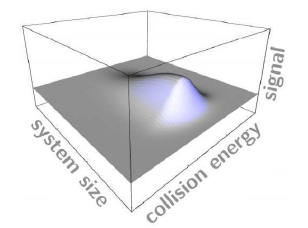}\hfill
 
  \caption{
    Hypothetical sketch of the phase diagram of strongly interacting matter with the critical point (CP), drawn as a function of baryochemical potential $\mu_{B}$ and temperature T (\textit{left}), experimental search for CP requires a two-dimensional scan in freeze-out
parameters (T, $\mu_{B}$) by changing collision energy and system size (\textit{left}, red). Theoretical studies predict the presence of a `hill of fluctuations' as a function of colliding system size and energy for observables sensitive to the CP ({\it right})
  }
  \label{scan}
\end{figure}

The experimental search for the critical point requires a two-dimensional scan in freeze-out parameters (T, $\mu_{B}$) by changing collision parameters controlled in the laboratory, i.e. energy and size of the colliding nuclei (or collision centrality) (Fig.~\ref{scan} {\it left}). A characteristic feature of a second-order phase transition (expected to occur at the CP) is the divergence of the correlation length. One expects a `hill' of increased fluctuations in various observables in the CP vicinity (Fig.~\ref{scan} {\it right}), see Ref.~\cite{MarekPeter} for more details.

\section{Experimental measures to search for critical point}
The expected signal of a critical point (CP) is a non-monotonic dependence of
various fluctuations (multiplicity, transverse momentum, baryon density).

\subsection{Extensive quantities}
An extensive quantity is a quantity that is proportional to the number of Wounded Nucleons ( {\it W} ) in
the Wounded Nucleon Model~\cite{Bialas:1976ed} (WNM) or volume ( {\it V} ) in the Ideal Boltzmann Grand
Canonical Ensemble (IB-GCE). The most popular is particle number (multiplicity) distribution $P(N)$
cumulants:
\begin{eqnarray*}
\kappa_{1} &=& \AVG{N},\\
\kappa_{2} &=& \AVG{(\delta N)^{2}} = \sigma^{2},\\
\kappa_{3} &=& \AVG{(\delta N)^{3}} = S\sigma^{3},\\
\kappa_{4} &=& \AVG{(\delta N)^{4}} - 3\AVG{(\delta N)^{2}}^{2} = \kappa\sigma^{4}.
\end{eqnarray*}
These multiplicity cumulants characterize the shape of multiplicity distribution and quantify fluctuations.


\subsection{Intensive quantities}

The ratio of any two extensive quantities is independent of \textit{W}~(WNM) or \textit{V}~(IB-GCE) for an event sample
with fixed \textit{W} (or \textit{V}) -- it is an intensive quantity. For example:
$$
  \AVG{A} / \AVG{B} = W\cdot\AVG{a} / W\cdot\AVG{b} = \AVG{a} / \AVG{b},
$$
where $A$ and $B$ are any extensive event quantities, i.e. $\AVG{A} \sim W$, $\AVG{B} \sim W$
and $\AVG{a} = \AVG{A}$ and $\AVG{b} = \AVG{B}$ for $W = 1$.\\
Popular examples are:
\begin{eqnarray*}
\kappa_{2}/\kappa_{1} &=& \omega[N] = \frac{\sigma^{2}[N]}{\AVG{N}} = \frac{W \cdot \sigma^{2}[n]}{W\cdot\AVG{n}} = \omega[n]\;\textrm{(scaled variance)},\\
\kappa_{3}/\kappa_{2} &=& S\sigma,\\
\kappa_{4}/\kappa_{2} &=& \kappa\sigma^{2}.
\end{eqnarray*}

\subsection{Strongly intensive quantities}

For an event sample with varying \textit{W} (or \textit{V}), cumulants are not extensive quantities anymore. For example:
$$
  \kappa_{2} = \sigma^{2}[N] = \sigma^{2}[n]\AVG{W} + \AVG{n}^{2}\sigma^{2}[W].
$$
However, having two extensive event quantities, one can construct quantities independent of
the fluctuations of \textit{W} (or \textit{V}). Popular examples include~\mbox{\cite{Gorenstein:2011vq,Gazdzicki:2013ana}}:
\begin{eqnarray*}
\AVG{K}/\AVG{\pi}, &&\\
\Delta[N,P_{T}] &=& (\omega[N]\AVG{P_{T}} - \omega[P_{T}]\AVG{N})/c,\\
\Sigma[N,P_{T}] &=& (\omega[N]\AVG{P_{T}} + \omega[B]\AVG{N} - 2(\AVG{NP_{T}} - \AVG{P_{T}}\AVG{N})/c,
\end{eqnarray*}
where $P_{T} = \sum\limits_{i=1}^{N} p_{T,i}$ and $c$ is any extensive quantity (e.g. $\AVG{N}$).

\subsection{Fluctuations as a function of momentum bin size}
When a system crosses the second-order phase transition, it becomes scale-invariant, which leads to a power-law form of the correlation function. This phenomenon leads to enhanced multiplicity fluctuations with special properties that can be revealed by scaled factorial moments (SFMs). The SFMs are calculated by partitioning a region of transverse momentum space into a lattice of \textit{M$\times$M} bins of equal size:
$$
          F_{r}(M) = \frac{\left\langle\frac{1}{M^{2}}\sum\limits_{i=1}^{M^{2}} n_{i}(n_{i}-1)...(n_{i} - r +1)\right\rangle}
							                {\left\langle\frac{1}{M^{2}}\sum\limits_{i=1}^{M^{2}} n_{i}\right\rangle^{r}},
$$
where $n_{i}$ is the number of particles in $i$-th bin, $M^{2}$ is the total number of bins, and averaging is done over bins and events ($\AVG{..}$).

In the case of a system exhibiting critical fluctuations, $F_{r}(M)$ is
expected to exhibit a power-law dependence on $M$~\cite{Wosiek:APPB,Bialas:1990xd,Bialas:1985jb,Antoniou:2006zb,Shuryak:2019acz}
$$
F_{r}(M) \sim M^{D\cdot\phi_{r}},
$$
where $D$ is the embedding dimension (for the transverse plane, $D=2$). Also, the exponent (intermittency index), $\phi_{r}$ is expected to obey the relation:
$$
D\cdot\phi_{r} = (r -1)\cdot d_{r},
$$
where the anomalous fractal dimension $d_{r}$ is independent of $r$.

\subsection{Scaled factorial moments in the Wounded Nucleon Model}
In the Wounded Nucleon Model, $N = \sum\limits_{i=1}^{W}n_{i}$, where $W$ is number of wounded nucleons (constant). Also $\AVG{N} = W\cdot\AVG{N}$ and $\omega[N] = \omega[n]$ (scale variance). The Second Scaled Factorial moment becomes:
$$
   F_{2}[N] = \frac{1}{W}F_{2}[n] + 1 - \frac{1}{W}.
$$
$F_{2}[N]$ is neither extensive ($\approx W$), nor intensive but in the limit $F_{2}[n]\gg 1$, $W\gg$ 1:
$$
  F_{2}[N] = \frac{1}{W}F_{2}[n],
$$
Scaled Factorial Moment is inversely extensive.
\section{Experimental results} 

\subsection{Multiplicity fluctuations}

\begin{figure}[!htb]
  \centering
  \vspace{0pt}\includegraphics[width=.35\textwidth]{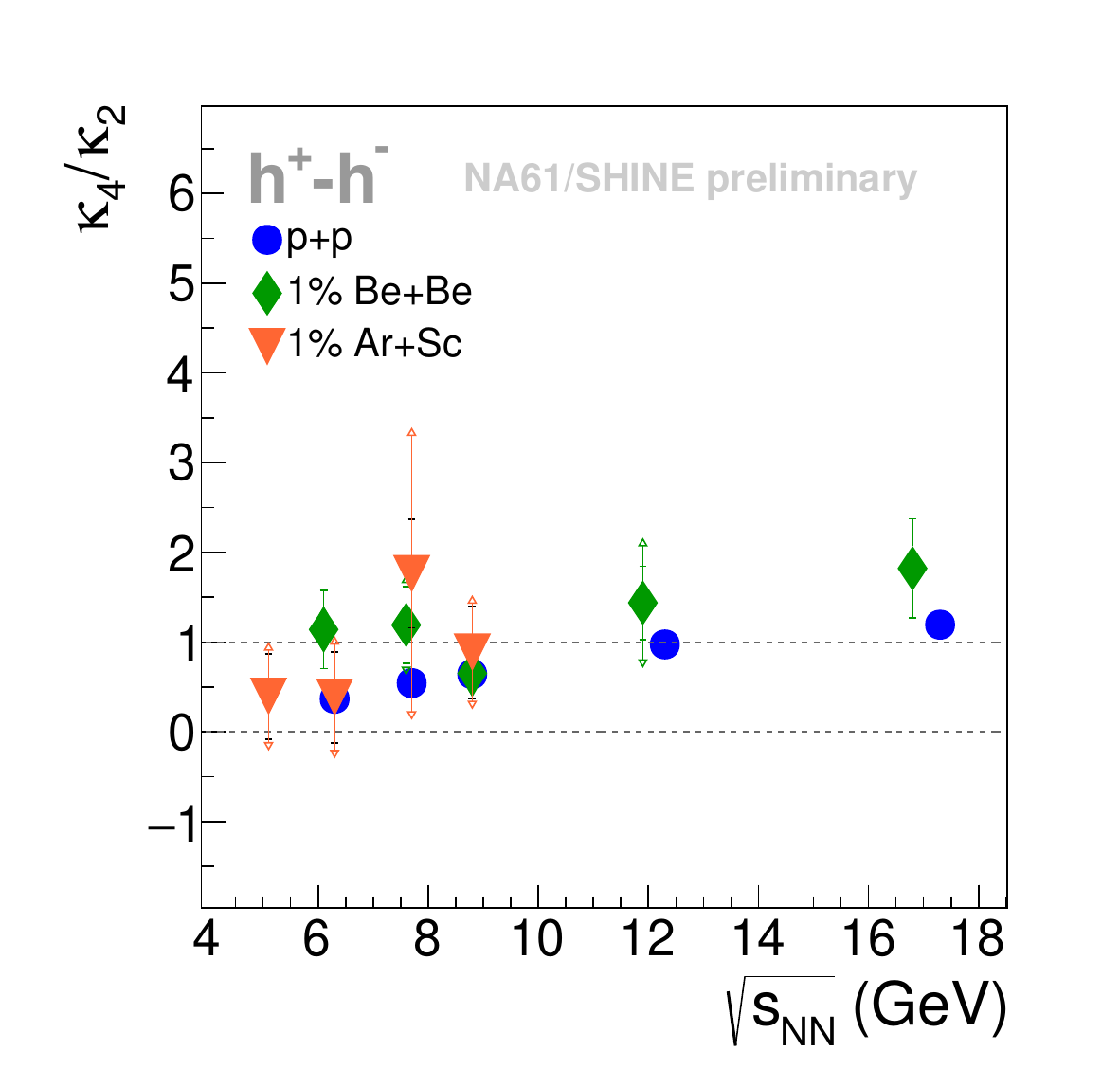}\hfill
  \vspace{0pt}\includegraphics[width=.48\textwidth]{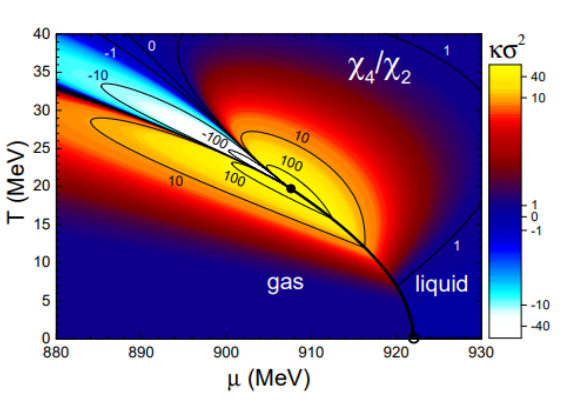}\hfill
   \caption{
  $\kappa_{4}/\kappa_{2}$ measured by NA61/SHINE~\cite{Gazdzicki:2017zrq} in inelastic \textit{p+p} interactions and \textit{central} Be+Be, Ar+Sc collisions at the CERN SPS energies.
  Critical fluctuations in model with Van der Waals interaction~\cite{MIGorennstein} ({\it right})
  }
  \label{fig:mult-fluct}
\end{figure}

 Results on the energy dependence of multiplicity fluctuations by NA61/SHINE~\cite{Gazdzicki:2017zrq} quantified by $\kappa_{4}/\kappa_{2}$ are presented in Fig.~\ref{fig:mult-fluct} ({\it left}).
No prominent structures that could be related to the critical point are observed.

\subsection{Multiplicity-transverse momentum fluctuations}
\begin{figure}[!htb]
  \centering
  \vspace{0pt}\includegraphics[width=.35\textwidth]{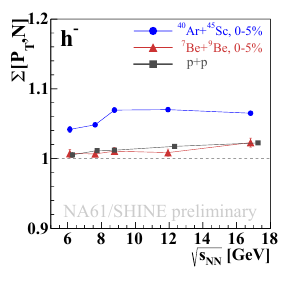}\hfill
  \vspace{0pt}\includegraphics[width=.5\textwidth]{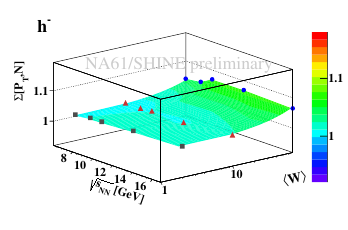}\hfill
 
  \caption{$\Sigma[P_{T},N]$ measured by NA61/SHINE in inelastic \textit{p+p} interactions and violent Be+Be, Ar+Sc collisions at the CERN SPS energies. Results refer to negatively charged hadrons at forward rapidity ($0 < y_{\pi} < y_{beam}$) and $p_{T} < 1.5~\GeVc$.
  }
  \label{fig:mult-pt-fluct}
\end{figure}

Results on the energy dependence of multiplicity-transverse momentum fluctuations by NA61/SHINE~\cite{Andronov:2016ddd} expressed in $\Sigma [P_{T},N]$, strongly intensive quantity, are presented in Fig.~\ref{fig:mult-pt-fluct}.
No prominent structures that could be attributed to the critical point are observed.

\subsection{Fluctuations as a function of momentum bin size (intermittency analysis)}
 The second Scaled Factorial Moment in transverse momentum space of mid-rapidity protons has been studied. The methodologies of using independent data points as well as cumulative quantities have been applied, as will be explained in detail below. Also, intermittency analysis of negatively charged hadrons ($h^{-}$) with Scaled Factorial Moments up to the 4th order have been studied.
\subsubsection{Cumulative transformation:}
Scaled Factorial Moments strongly depend on the single-particle transverse momentum distribution. There are two possible ways to minimize it. First, construct mixed event data set by mixing particles from the recorded data such that each mixed event consists of particles from different data events. Then the scaled-factorial moment of mixed data should be subtracted from the recorded data. The other possibility is to transform transverse momentum components, $p_{x}$ and $p_{y}$, into their cumulative equivalents~\cite{BilasMarek}. This transforms non-uniform distributions into uniform from 0 to 1, preserving.
(approximately) the power-law relation~\cite{Diakonos}.
\begin{figure}[!htb]
    \centering
    \includegraphics[width=5.5in]{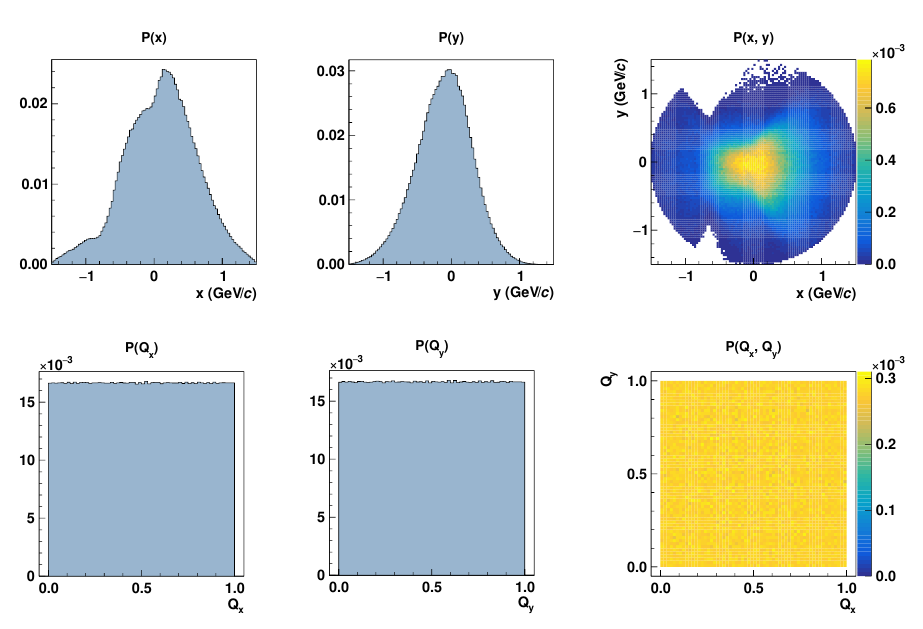}
    \caption{Example of the effect of the cumulative transformation of transverse momentum components, $p_{x}$ and $p_{y}$, of proton candidates selected for intermittency analysis of NA61/SHINE Ar+Sc at 150\AGeVc~ data.
Distributions before (\textit{top}) and after (\textit{bottom}) the transformation are presented.}
    \label{cumulative}
\end{figure}

An example of the effect of the transformation is shown in Fig.~\ref{cumulative}. Initial transverse momentum components, $x$ and $y$, have been transformed into their cumulative equivalents, $Q_{x}$ and $Q_{y}$. As a result, their non-uniform distributions became uniform between 0 and 1.

\subsubsection{Proton intermittency results: }
\begin{figure}[!htb]
    \centering
    \includegraphics[width=5in]{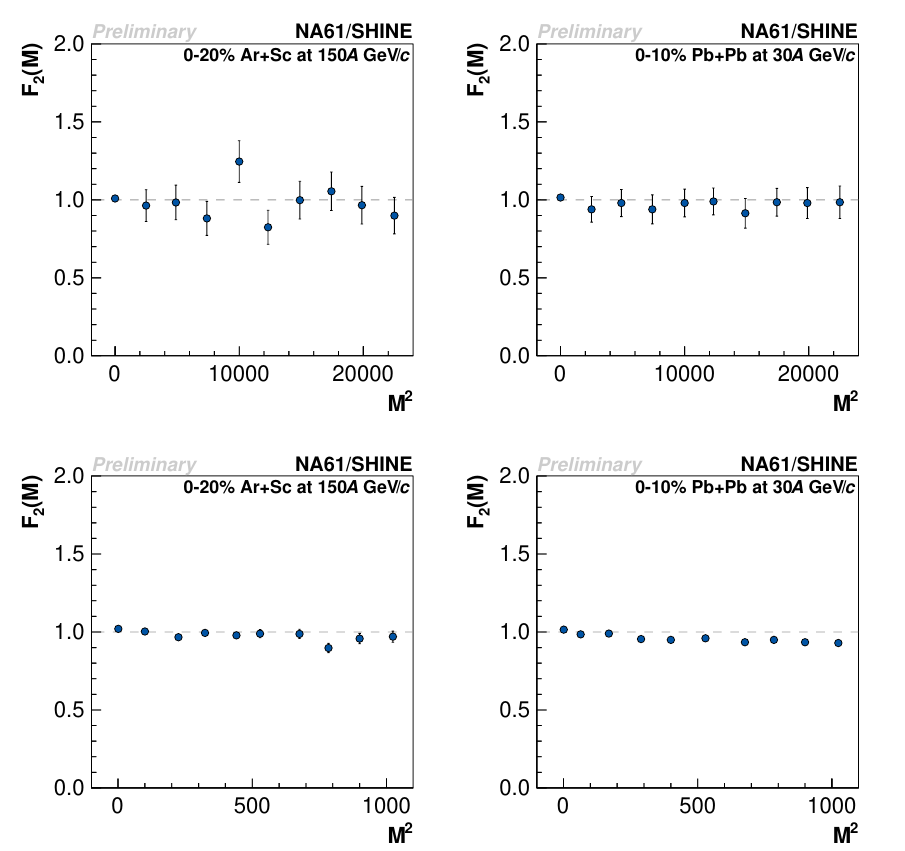}
    \caption{Results on the dependence of the second scaled factorial moment of the mid-rapidity proton
multiplicity distributions for Ar+Sc at 150\AGeVc~ (\textit{left}) and Pb+Pb at 30\AGeVc~ (\textit{right}). Results were obtained
for M from 1 to 150 (\textit{bottom}) and 1 to 32 (\textit{top}). Only statistical uncertainties are shown.} 
\label{result}
\end{figure}

In Fig.~\ref{result}, results on the dependence of the Second Scaled Factorial Moments of proton multiplicity distributions in mid-rapidity for 0-20\% most central inelastic Ar+Sc at 150\AGeVc ($\sqrt{s_{NN}} \approx 17$ GeV) and 0-10\% most central Pb+Pb at 30\AGeVc ($\sqrt{s_{NN}} \approx 7.5$ GeV) collisions are presented.
Each of the ten points is calculated using a fraction of the total available statistics (independent points), and cumulative transverse momentum components were used.\\
A separate study~\cite{TobiaszSubhasis} has shown that the momentum resolution of the detector may substantially distort the power-law form of $F_{2}(M)$ for very small bin sizes. Therefore, $F_{2}(M)$ was analyzed in two ranges of momentum
bin sizes, i.e. from 1 to 150 (Fig.~\ref{result}, \textit{top}) and from 1 to 32 (Fig.~\ref{result}, \textit{bottom}).
No indication of a power-law increase in the number of bins is observed.

\subsubsection{Comparison with a power-law model:}
Results are compared with a power-law model~\cite{Subhasis}. It generates momenta of uncorrelated and correlated protons in events with a given multiplicity distribution. The model has two controllable parameters: the fraction of correlated particles and the strength of the correlation (the power-law exponent).
Using the power-law model, many high statistics data sets with multiplicity distributions identical to the experimental data and similar inclusive transverse momentum distributions have been produced.
Each data set has a different fraction of correlated particles (varying from 0 to 4\%) and/or a different power-law exponent (varying from 0.0 to 1.0).
\begin{figure}[!htb]
    \centering
    \includegraphics[width=5.3in]{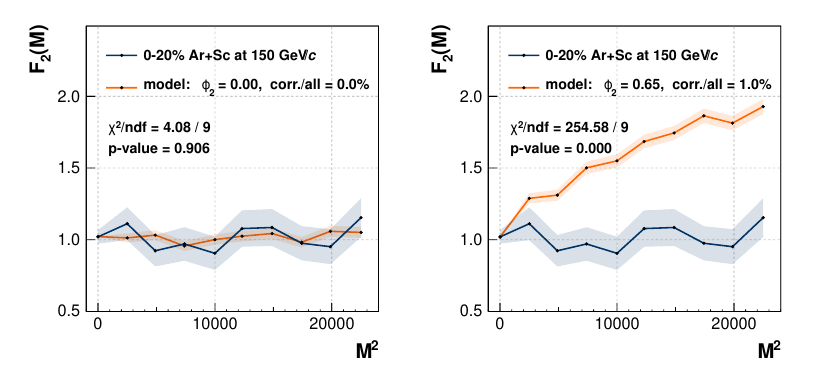}
    \caption{Example of the comparison of the simple power-law model data with experimental Ar+Sc at 150\AGeVc. No correlated particles (\textit{left}) and 1\% particles correlated with a power $\phi_{2}$ = 0.65 (right). Obtained $\chi^{2}$ and p-value are used to construct the exclusion plot} 
\label{toymodel}
\end{figure}
Next, all these generated data sets have been analyzed in the same way as the experimental data. $F_{2}(M)$ results have been compared, and the p-value (probability of making a mistake by rejecting model parameters which do not reproduce the data) was calculated. An example of such comparisons is presented in Fig.~\ref{toymodel}. 
\begin{figure}[!htb]
    \centering
    \includegraphics[width=5.3in]{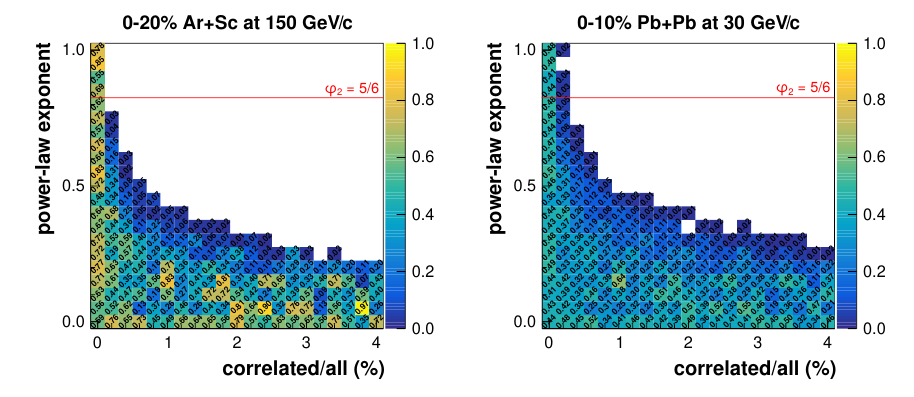}
    \caption{Exclusion plots: p-values for comparisons of $F_{2}$(M) for Ar+Sc at 150\AGeVc~ (\textit{left}) and Pb+Pb at 30\AGeVc~(\textit{right}) data with simple power-law model data sets with various values of two parameters (fraction of correlated particles and power-law exponent). White areas correspond to the p-value of less than 1\%.
Theoretical prediction for the power value for a system freezing out at the QCD critical point is marked with
the red line.} 
\label{Exclusionplot}
\end{figure}
All p-values from each model data set, presented in one plot, form an exclusion plot (Fig.~\ref{Exclusionplot}).
The color scale is related to the probability that the model data set for a given number of correlated particles and a given power might be describing the data. White areas correspond to the p-value of less
than 1\% and may be considered excluded (for this particular model). The intermittency index $\phi_{2}$ for a system freezing out at the QCD critical point is expected to be $\phi_{2} = 5/6$, assuming that the latter belongs to the 3-D Ising universality class.
\subsubsection{Negatively charged hadron(h$^{-}$) intermittency result: }

\begin{figure}[!htb]
    \centering
    \includegraphics[width=3.2in]{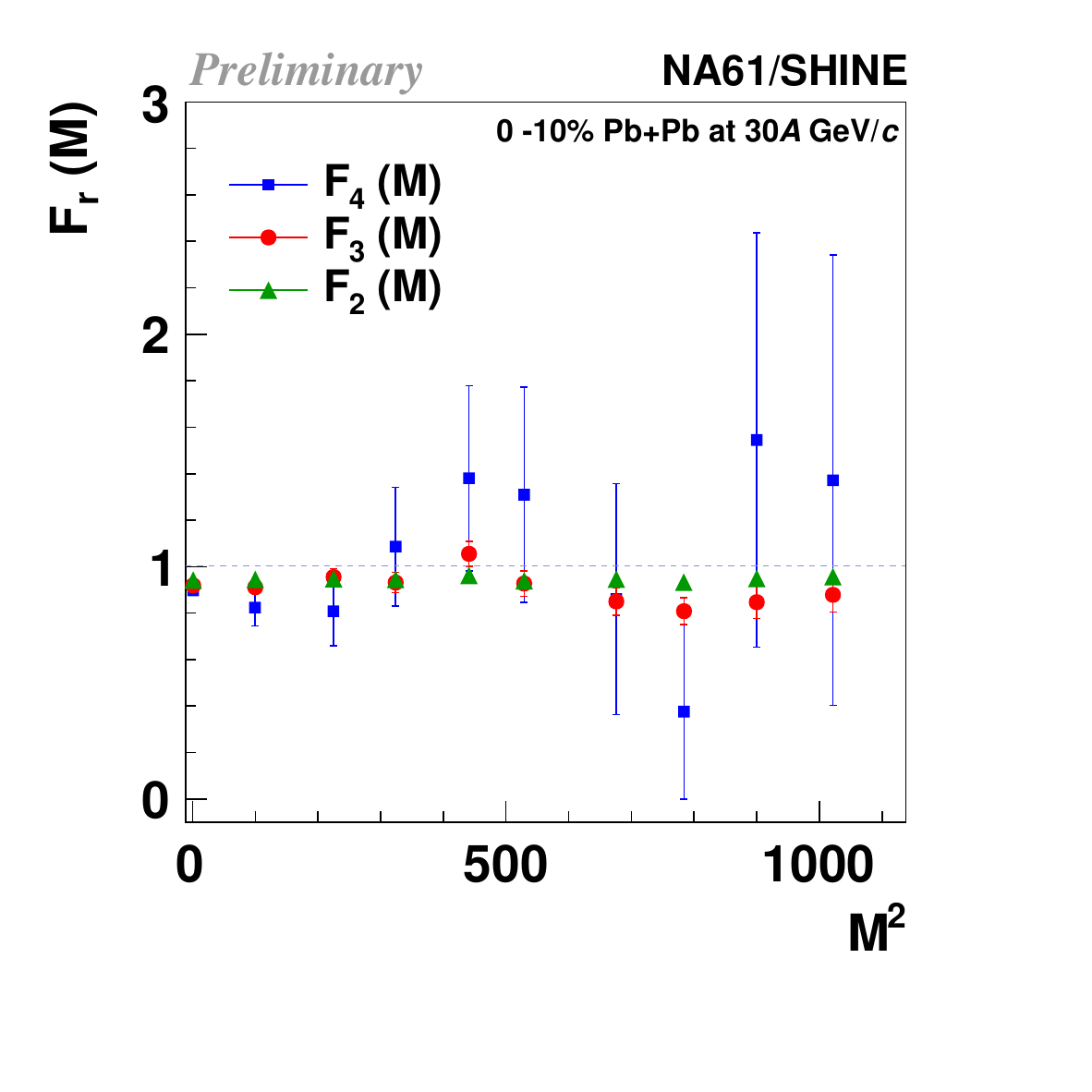}
    \caption{Results on the scaled factorial moments' dependence on the mid-rapidity h$^{-}$ multiplicity distributions for  Pb+Pb at 30\AGeVc. The result obtained
for M from 1 to 32 up to 4th order scaled factorial moments. Only statistical uncertainties are shown.} 
    
    \label{pion}
\end{figure}

In Fig.~\ref{pion}, $h^{-}$ result on the dependence of the Scaled Factorial Moments up to 4th order of mid-rapidity $h^{-}$ multiplicity distributions for 0-10\% most central Pb+Pb at 30\AGeVc collisions. Each of the ten points is calculated using a fraction of the total available statistics, and cumulative transverse momentum components were used.
No indication of a power-law increase in the number of bins is observed.
\section{Summary}
NA61/SHINE experimental results on QCD critical point search have been presented. No prominent structures that could be attributed to the QCD critical point are observed in the results on multiplicity and multiplicity-transverse momentum fluctuations. No power-law increase was observed via proton and negatively charged hadron intermittency analysis of Pb+Pb at 30\AGeVc and proton intermittency analysis of Ar+Sc at 150\AGeVc data. The exclusion plot for the parameters of the power-law model (Fig.~\ref{Exclusionplot}) for Ar+Sc at 150\AGeVc and Pb+Pb at 30\AGeVc was obtained.

\section{Acknowledgments}
This work is supported by the National Science Centre, Poland, under grant no. 2018/30/A/ST2/0026.

\end{document}